\documentclass[prl,twocolumn,showpacs]{revtex4}
\usepackage{graphicx}
\usepackage[ansinew]{inputenc}
\usepackage[english]{babel}
\usepackage{pst-all}

\DeclareGraphicsExtensions{.pdf,.ps,.eps}

\begin{document}

\title{Effects of interactions in transport through 
Aharonov-Bohm-Casher interferometers}
\author{A. M. Lobos}
\author{A. A. Aligia}
\affiliation{Centro At\'omico Bariloche and Instituto Balseiro, Comisi\'on Nacional de
Energ\'{\i}a At\'omica, 8400 Bariloche, Argentina}
\date{\today}

\begin{abstract}
We study the conductance through a ring described by the Hubbard model (such
as an array of quantum dots), threaded by a magnetic flux and subject to
Rashba spin-orbit coupling (SOC). We develop a formalism that is able to
describe the interference effects as well as the Kondo effect when the
number of electrons in the ring is odd. In the Kondo regime, the SOC reduces
the conductance from the unitary limit, and in combination with the magnetic
flux, the device acts as a spin polarizer.
\end{abstract}

\maketitle

Advances in semiconductor technology have provided useful tools to test
fundamental concepts of quantum physics, such as the superposition principle
and the existence of topological phases \cite{berry84}. Beautiful
demonstrations of these are studies of the Aharonov-Bohm (AB) effect 
\cite{aharonov&bohm59} in mesoscopic rings, particularly with embedded 
quantum dots (QDs) \cite{ji00,wiel00}. 
The effect of interactions
in these systems is still a matter of debate \cite{jiang04}. Despite the
enormous effort to describe transport through interacting regions \cite{meir92}, at present we do not have a unified procedure to extend the
results of the single particle case to many-body cases. A serious
shortcoming of the calculation of the conductance $G$ through an interacting
ring is that even knowing the exact eigenstates of the ring, there is no
simple procedure to calculate $G$. When the coupling $V$ of the
ring to the conducting leads is small, Jagla and Balseiro (JB) used a
perturbative expression in $V$ for $G$ that is exact for any $V$
in the non-interacting limit \cite{jagla93}. Similar equations were used
recently, assuming that a Zeeman term destroys the Kondo effect in the
system \cite{hallberg04,aligia04}. Another expression in order $V^{2}$ was
proposed last year \cite{pletyukhov06}. Unfortunately these expressions are
not valid in the Kondo regime, in which the number of electrons in the ring
is odd, because the resulting Kondo physics cannot be described by
perturbation theory in $V$. The ideal conductance in the
Kondo regime was recovered by mapping the model into an impurity Anderson
model, but in this formulation interference effects were 
lost \cite{aligia04}.

Recently, the Aharonov-Casher (AC) effect \cite{aharonov&casher84}, the
charge-spin dual of the AB effect, has been demonstrated experimentally in
semiconductor mesoscopic rings \cite{konig06, bergsten06}. The AC phases are
originated due to the Rashba spin-orbit coupling (SOC) in the ring,
resulting from electronic motion in the presence of an electric field normal
to the plane of the ring. The interference between electrons of given spin
travelling clockwise and anticlockwise produces a strong modulation of the
electronic current through the device. Recent theoretical research 
\cite{molnar04} has successfully explained the modulation
of the conductance in terms of non-interacting electrons. However, the
single-electron picture, turns out
to be inadequate to describe electronic transport in the
strongly-interacting case, particularly in the Kondo regime, as we will show.

In this Letter, we describe a systematic procedure to
calculate equilibrium conductance $G$ through a ring of an interacting system
weakly coupled to conducting leads, that takes into account both the effects
of interference and correlations in presence of a magnetic flux and SOC.
Using a non-abelian gauge transformation (NAGT), we show that for on-site
interactions, the SOC can be absorbed in opposite AC phases for spin up and
down in an adequately chosen quantization axis. For a Hubbard model (that
describes a ring of an even number of QDs),
in absence of SOC $G$ vanishes when the magnetic flux amounts to half a flux quantum.
For other fluxes in the Kondo regime, $G$ reaches the unitary
limit (ideal conductance \cite{wiel00}). When the SOC is turned on, the 
ideal conductance is destroyed and $G$ shows a strong spin dependence in this
regime.

Our first task is to derive the appropriate extension to the Hubbard model
to include the SOC in an adequate representation that simplifies our
subsequent calculations. To illustrate the procedure, it is easier to begin
with non-interacting electrons in the continuum. The correct
Hamiltonian for this case was derived by Meijer 
\textit{et al.} \cite{meijer02}. The SOC
is $H_{\text{SOC}}=\alpha ^{\prime }\vec{\sigma}\cdot \vec{\mathrm{E}}\times
(\vec{\mathrm{p}}-e\vec{\mathrm{A}})$, where
$\alpha ^{\prime }$ is the Rashba constant and
$\vec{\mathrm{E}}$ is the
electric field, which in our case is in the $z$ direction, perpendicular to
the plane of the ring. Including SOC and the orbital effects of the magnetic
field, but neglecting the Zeeman term (usually several orders of magnitude
smaller than the Kondo energy scale in QDs \cite{wiel00}), 
the Hamiltonian can be written in the form \cite{molnar04} (a)
\begin{equation}
H_{\text{NI}}=\hbar \Omega \left[ -i\frac{\partial }{\partial \varphi }-%
\frac{\phi }{\phi _{0}}+\frac{\omega _{so}}{2\Omega }\sigma _{r}(\varphi )%
\right] ^{2},  \label{hnoninteracting}
\end{equation}%
where $\Omega =\hbar /(2m^{\ast }r^{2})$, $m^{\ast }$ is the effective
electron mass, $r$ is the radius of the ring, $\omega _{so}=\alpha /\hbar r$%
, $\alpha =\hbar \alpha ^{\prime }E_{z}$, $\phi =B\pi r^{2}$ is the magnetic
flux, $\phi _{0}=h/e$ is the flux quantum and $\sigma _{r}(\varphi )=$ $%
\sigma _{x}\cos {\varphi }$ + $\sigma _{y}\sin {\varphi }$ is the Pauli
matrix in the radial direction,
and $\varphi$ is the azimuthal angle (see Fig. 1).
Although the Schr\"odinger equation 
$H_{\text{NI}}\chi (\varphi )=E\chi (\varphi )$ (where $\chi $ is a spinor) has been
solved \cite{molnar04}, we are interested in a simplification of this
equation that can be extended to the interacting case. This can be achieved
by a NAGT $\chi (\varphi )=\hat{U}(\varphi )\chi ^{\prime }(\varphi )$,
where the  operator $\hat{U}(\varphi )$ satisfies the differential equation

\begin{equation}
i\frac{\partial }{\partial \varphi }\hat{U}(\varphi )=\left[ -\frac{\phi }{%
\phi _{0}}+\frac{\omega _{so}}{2\Omega }\sigma _{r}(\varphi )\right] \hat{U}%
(\varphi ).  \label{equ}
\end{equation}
It can be easily checked that in the transformed Hamiltonian, $H_{\text{NI}%
}^{\prime }=\hat{U}^\dagger H_{\text{NI}}\hat{U}$ $=-\hbar \Omega \partial
^{2}/\varphi ^{2}$ the magnetic flux and the SOC disappeared, and enter now
in the boundary condition, since $\chi (2\pi )=\chi (0)$ implies $\chi
^{\prime }(2\pi )=\hat{U}^{\dagger }(2\pi )\chi ^{\prime }(0)$. The solution
of Eq.(\ref{equ}) with $\hat{U}(0)=1$ is

\begin{equation}
\hat{U}(\varphi )=\mathrm{exp}\left[ -i\sigma _{z}\frac{\varphi }{2}\right] 
\mathrm{exp}\left[ i\vec{\sigma}.\vec{n}_{\theta }
\frac{\varphi ^{\prime }}{2}\right] \mathrm{exp}
\left[ i\frac{\phi }{\phi _{0}}\varphi \right] ,
\label{u}
\end{equation}
where $\vec{n}_{\theta }=(-\sin {\theta },0,\cos {\theta })$, $\theta
=\arctan {(\omega _{so}/\Omega )}$, and $\varphi ^{\prime }=\varphi 
\sqrt{1+(\omega _{so}/\Omega )^{2}}$.

To construct the tight binding version of $H_{\text{NI}}^{\prime }$, 
let us assume that we have $N$ sites, lattice parameter $a$ (with $%
Na=2\pi r$) and site 0 at angle $\varphi =0$. For simplicity we consider
only hopping between nearest neighbors (NN). Then, we can take a constant
hopping $t$ between all NN, except between sites $N-1$ and 0, in which the
boundary condition should be included. The matrix $\hat{U}(2\pi )$ is easily
diagonalized in the quantization axis $\vec{n}_{\theta }$ and its
eigenvalues are $\exp [i(\Phi _{AB}+\sigma \Phi _{AC})]$, where $\sigma =1$
(-1) for spin up (down) in this direction, $\Phi _{AB}=2\pi \phi /\phi _{0}$, 
and $\Phi _{AC}=\pi \{ [1+(\omega _{so}/\Omega )^{2}]^{1/2}-1 \}$. Therefore,
destroying a particle with spin $\sigma $ at site $N-1$ and creating it at
site 0 should be accompanied by the corresponding exponential factors. 
On-site interactions are not affected by the NAGT.
Changing the phases of the second quantization operators to write the 
Hamiltonian in rotationally invariant form,
the transformed Hubbard model in the ring becomes

\begin{eqnarray}
H_{r}^{\prime }&=&\sum_{i=0,\sigma }^{N-1}t\left[ e^{i(\Phi _{AB}+\sigma
\Phi _{AC})/N}d_{i+1\sigma }^{\dagger }d_{i\sigma }+\mathrm{H.c.}\right] + 
\nonumber \\
&&+Ud_{i\uparrow }^{\dagger }d_{i\uparrow }d_{i\downarrow }^{\dagger
}d_{i\downarrow }.  \label{hub}
\end{eqnarray}
From the curvature of the dispersion relation at small wave vector 
$t=\hbar^2 /(2m^{*}a^{2})$, and then $\omega _{so}/\Omega =\alpha N/(2\pi ta)$.
Thus, the AC phase can be written as

\begin{equation}
\frac{\Phi _{AC}}{N}=\sqrt{\left( \frac{\pi }{N}\right)
^{2}+\left( \frac{\alpha }{2ta}\right) ^{2}}-\frac{\pi }{N}.  \label{ac}
\end{equation}%
Therefore, for large $\alpha $ or $N$, the properties of the system are
periodic with $\alpha $ as observed experimentally \cite{konig06, bergsten06}.

The fact that the SOC can be gauged away in one dimension has been noted
previously \cite{meir89}, but the explicit form of the transformation has
not been derived. This transformation has important consequences. In the
thermodynamic limit the boundary conditions are irrelevant and therefore the
thermodynamic properties of the system should be identical to those of the
Hubbard model without SOC. This is not obvious in alternative treatments 
\cite{gritsev05}. In particular it seems that the opening of a spin gap in
the system requires long-range interactions.

\begin{figure}[ht]
\includegraphics[scale=0.33,clip]{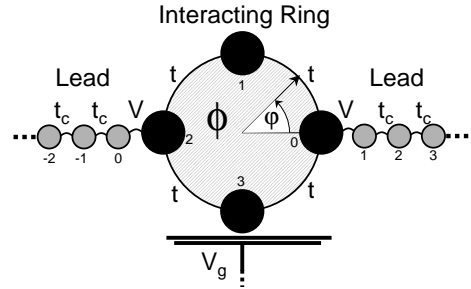}
\caption{Scheme of the system.}
\label{system}
\end{figure}

To study the conductance, we must consider the Hamiltonian of the complete
system $H=H_{l}+H_{r}^{\prime }+H_{V}$, where with the appropriate
quantization axis \cite{note_homomorphism} and choice of phases 
$H_{l}=t_{c}(\sum_{i=0,\sigma}^{-\infty }c_{i-1,\sigma }^{\dagger }c_{i\sigma }$ 
+ $\sum_{i=1,\sigma }^{\infty }c_{i+1,\sigma }^{\dagger }c_{i\sigma }$ + 
H.c.), describes the non-interacting leads, and $H_{V}=V(\sum_{%
\sigma }c_{0\sigma }^{\dagger }d_{N/2,\sigma }$ + $c_{1\sigma }^{\dagger
}d_{0\sigma }$ + \textrm{H.c.)} is the coupling between the ring and leads.
For simplicity we will focus here on the particular case $N=4$, illustrated
in Fig. \ref{system}. 
We assume that the leads are described by a constant density of states
$\rho_0=1/W$, and we take for the band width of the leads $W=60 t$ 
($W$ is usually much larger than $t$ in QD arrays).
The Fermi level is set 
at $\epsilon _{F}=0$. 
To control the charge in the ring we add to $H_{r}^{\prime
}$ a term $-V_{g}\sum_{i\sigma }d_{i\sigma }^{\dagger }d_{i\sigma }$ that
represents the effect of a gate voltage. Our approximations to calculate $G$
amount to a truncation of the Hilbert space of $H_{r}^{\prime }$
and a slave boson mean-field approximation for the resulting
generalized Anderson model (GAM). $H_{r}^{\prime }$ can be diagonalized
exactly (numerically for not too large $N$). We retain only two neighboring
charge configurations with $n$ and $n-1$ particles and we have chosen $n=4$.
Furthermore, we retain only the lowest lying singlet state for 4 particles 
($|\psi _{0}^{4}\rangle $ with energy $E_{0}^{4}$) and all doublets for 3
particles. This procedure is valid for small enough $V$ 
\cite{lobos06}. Calculating the matrix elements of $H_{V}$ in the truncated
Hilbert space leads to a GAM 
\begin{eqnarray}
H_{\text{GAM}} &=&H_{l}+\sum_{j,\sigma} E_{j\sigma }^{3}|\psi _{j\sigma }^{3}\rangle
\langle \psi _{j\sigma }^{3}|+E_{0}^{4}|\psi _{0}^{4}\rangle \langle \psi
_{0}^{4}|+  \nonumber \\
&&+V\sum_{ j, \sigma, \eta=0,1} (\alpha _{j\sigma }^{\eta }|\psi _{0}^{4}\rangle \langle \psi
_{j\sigma }^{3}|c_{\eta \sigma }+\mathrm{H.c.)},  \label{ham}
\end{eqnarray}
where $|\psi _{j\sigma }^{3}\rangle $ and $E_{j\sigma }^{3}$ denote the 
$j$-th eigenvector and eigenvalue of $H_{r}^{\prime }$ in the configuration
with 3 particles with spin projection $\sigma $, in ascending order of
energy and 
\begin{eqnarray}
\alpha _{j\sigma }^{1} =\langle \psi _{0}^{4}|d_{0\sigma }^{\dagger }|\psi
_{j\overline{\sigma }}^{3}\rangle ,  
\alpha _{j\sigma }^{0} =\langle \psi _{0}^{4}|d_{N/2,\sigma }^{\dagger
}|\psi _{j\overline{\sigma }}^{3}\rangle .  \label{alpha}
\end{eqnarray}
$H_{\text{GAM}}$ can be expressed exactly in terms of a slave-boson
representation similar to that proposed by Coleman \cite{coleman86}: $|\Psi
_{j\sigma }^{3}\rangle \langle \Psi _{j\sigma }^{3}|\rightarrow f_{j\sigma
}^{\dagger }f_{j\sigma }$, $\ |\Psi _{0}^{4}\rangle \langle \Psi
_{0}^{4}|\rightarrow b^{\dagger }b$, $|\Psi _{j\sigma }^{3}\rangle \langle
\Psi _{0}^{4}|\rightarrow f_{j\sigma }^{\dagger }b$ and $|\Psi
_{0}^{4}\rangle \langle \Psi _{j\sigma }^{3}|\rightarrow b^{\dagger
}f_{j\sigma }$, where the operators $b^{\dagger }$ and $f_{j\sigma
}^{\dagger }$ create a boson and a fermion respectively and are subject to
the constraint $\sum_{j\sigma }f_{j\sigma }^{\dagger }f_{j\sigma
}+b^{\dagger }b=1$, which is incorporated in the Hamiltonian with a Lagrange
multiplier $\lambda $. We perform a saddle-point approximation in the
bosonic degrees of freedom, which reproduces the Kondo physics at low
temperatures 
and becomes exact in the limit of infinite degeneracy of the magnetic configuration, 
due to vanishing fluctuations around the mean-field value of the bosonic field 
\cite{coleman86}. The problem becomes equivalent to an effective
non-interacting fermionic Hamiltonian, with parameters $b_{0},\lambda$
(where $b_{0}=\langle b^{\dagger }\rangle =\langle b\rangle $) which are
determined by minimization of the free energy. Thus, we can use the
two-terminal Landauer formula to calculate the conductance, giving at
zero temperature \cite{meir92}
\begin{eqnarray}
G &=&\sum_{\sigma }G_{\sigma }, \\
\frac{G_{\sigma }}{G_{0}} &=&2\left( \pi \rho _{0}V^{2}b_{0}^{2}\right)
^{2}|\sum_{i,j}\bar{\alpha}_{i\sigma }^{0}\alpha _{j\sigma
}^{1}g_{i,j}^{\sigma }(\epsilon _{F})|^{2},
\label{conductance}
\end{eqnarray}
where $G_{0}$ is $2e^{2}/h$ and 
\[
g_{i,j}^{\sigma }=g_{i,j}^{0\sigma }+\frac{b_{0}^{2}V^{2}g_{ii}^{0\sigma
}g_{jj}^{0\sigma }}{A_{11}^{\sigma }A_{00}^{\sigma }-A_{10}^{\sigma
}A_{01}^{\sigma }}\sum_{\eta ,\eta ^{\prime }}\alpha _{i\sigma }^{\eta }%
\overline{\alpha }_{j\sigma }^{\eta ^{\prime }}A_{\eta \eta ^{\prime
}}^{\sigma }
\]
where $g_{ij}^{0\sigma }\equiv g_{ij}^{0\sigma }(\omega )=\delta
_{ij}(\omega -E_{j\sigma }^{3}-\lambda )^{-1}$ is the propagator of the $j$-th
state of 3 particles with spin projection $\sigma $ in the isolated ring,
and the functions $A_{\eta \eta ^{\prime }}^{\sigma }$ are $A_{\eta \eta
}^{\sigma }=1-b_{0}^{2}V^{2}g_{\eta }^{(0)}\sum_{m}|\alpha _{m\sigma }^{\eta
}|^{2}/(\omega -E_{m}^{3}-\lambda )$ for $\eta =\eta ^{\prime }$ and $%
A_{\eta \eta ^{\prime }}^{\sigma }=b_{0}^{2}V^{2}g_{\eta
}^{(0)}\sum_{m}\alpha _{m\sigma }^{\eta }\overline{\alpha }_{m\sigma }^{\eta
^{\prime }}/(\omega -E_{m}^{3}-\lambda )$ for $\eta \neq \eta ^{\prime }$,
where $g_{\eta }^{(0)}(\omega )$ is the Green's function at site $\eta $ of
the corresponding isolated lead ($V=0$).

\begin{figure}[ht]
\centering \includegraphics[scale=1.7,clip,keepaspectratio]{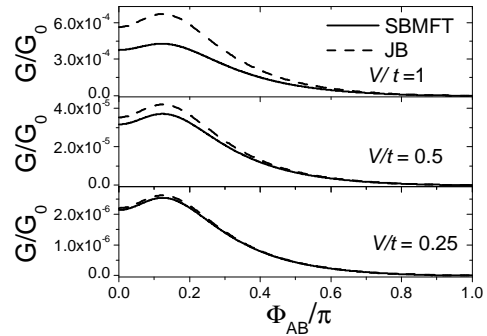}
\caption{Conductance as a function of magnetic flux for $\protect\alpha=0$, $%
V_g=0.8t $, $U=2t$ and several values of $V$. Full lines: our formalism. Dashed 
lines: JB expression.}
\label{nonmag}
\end{figure}
In Fig. \ref{nonmag} we show $G$ as a function of
magnetic flux in the non-magnetic regime 
$E_{0}^{(4)}<E_{0 \sigma}^{(3)}$, 
for different values of $V$ and without electric field ($\alpha =0$). The
conductance is even with flux \cite{note_invariance} and therefore, 
it is enough to show $G$
in the interval $0 \le \phi \le \phi _{0}/2$ (or $0 \le \Phi _{AB}\le \pi $). 
In this
regime correlations play a minor role and one expects that the JB formula 
\cite{jagla93}, which is exact in the non-interacting case, gives accurate
values for $G$. Our results show the same qualitative behavior.
In fact, for small $V$ it can be demonstrated that both approaches are
equivalent in this regime. For $\phi \le \phi _{0}/2$, $G$
vanishes due to destructive interference.

\begin{figure}[ht]
\centering 
\includegraphics[scale=2.0,clip,keepaspectratio]{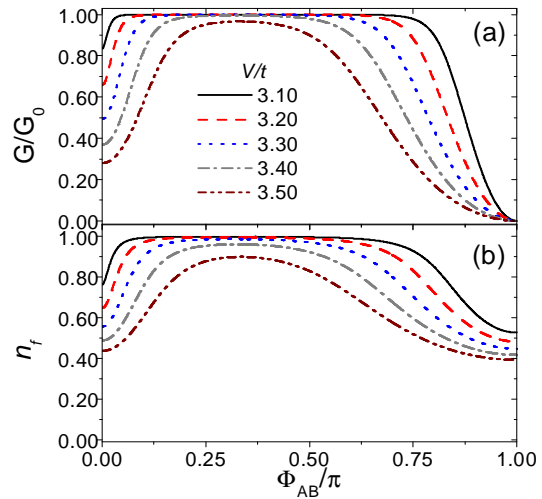}
\caption{(a) Conductance and (b) occupancy as a function of flux
for $\alpha =0$, 
$V_{g}=0$, $U=6t$ and several values of $V$.}
\label{kondo}
\end{figure}

The difference $E_{0}^{(4)}- E_{0 \sigma}^{(3)}$ can be reduced and turned 
negative applying a negative gate voltage.
The most important results of this work are those obtained in this case 
i.e., when the ring is in the mixed valence or Kondo
regime. Results for $\alpha =0$ are presented in Fig. \ref{kondo} (a). 
For small enough $\Delta /|E_{0}^{(3)}-E_{0}^{(4)}|$, where $\Delta =\pi
\rho _{0}V^{2}(|\alpha _{0\sigma }^{0}|^{2}+|\alpha _{0\sigma }^{1}|^{2})$,
charge fluctuations are frozen and a clear signature of Kondo physics is
displayed in the characteristic plateau in $G$ at the ideal conductance $%
G_{0}$ (the unitary limit) \cite{wiel00}. This is shown in the figure for the smaller
values of $V$ at small fluxes. The dependence of $G$ with flux, is related
with the corresponding dependence of the energy levels and matrix elements
with $\phi $. For larger $V$ and $\Phi _{AB}\sim \pi $, the system is in the
intermediate valence regime, as reflected in Fig. \ref{kondo} (b) in which
the total occupancy of the configuration with three particles $%
n_{f}=\sum_{j\sigma }\langle f_{\sigma }^{\dagger }f_{\sigma }\rangle $ is
shown. Therefore, the conductance deviates from the unitary limit.

Independently of the other parameters, $G$ vanishes at $\Phi _{AB}=\pi $. 
Within our formulation, at this point the states of the $n=3$ configuration
become doubly degenerate between states of different parity. The matrix
elements $\alpha _{j\sigma }^{\eta }$ entering Eq. (\ref{conductance}) have
the same modulus but different sign, therefore producing a complete
destructive interference inside the absolute value. To our knowledge, there
are no calculations so far showing at the same time this destructive
interference and ideal conductance in the Kondo regime. The JB expression
gives values below $0.1G_{0}$ for all $\Phi _{AB}$ and parameters of Fig. 
\ref{kondo}. Previous mappings to the Anderson model displayed the
Kondo physics, but did not capture the interference \cite{aligia04,lobos06}.

\begin{figure}[ht]
\centering \includegraphics[scale=0.8,clip,keepaspectratio]{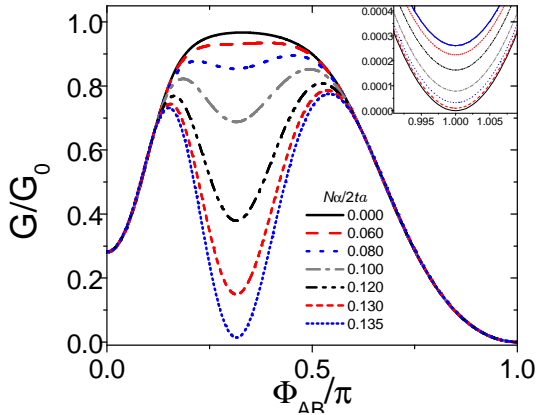}
\caption{Conductance as a function of flux for $V=3.5t$ and
different values of $\alpha$. Other parameters as in Fig. 
\protect\ref{kondo}.}
\label{rashba1}
\end{figure}

The effect of the SOC on the total conductance is dramatic in the Kondo
regime. The results presented in Fig. \ref{rashba1} show dips (additional to
that of $\Phi _{AB}=\pi $) which are larger as $\alpha $
grows. The main difference with the case $\alpha =0$ is that $n_{f\uparrow
}\neq n_{f\downarrow }$ therefore producing a partial destruction of the
Kondo resonance, mimicking the effect of a Zeeman term. This effect is
larger for lower $\Delta $ (when the system is deeper inside the Kondo
regime), which for the parameters of Fig. \ref{rashba1}, corresponds to 
$\Phi _{AB}\sim \pm 0.3\pi $. For $\Phi _{AB}=\pi $, complete
cancellation is not achieved. This can be interpreted as an effect of
quasiparticles acquiring relative phases differing slightly from $\pi $ due
to the effect of the additional AC phase (see inset in Fig. \ref{kondo}).

\begin{figure}[ht]
\centering \includegraphics[scale=2.0,clip,keepaspectratio]{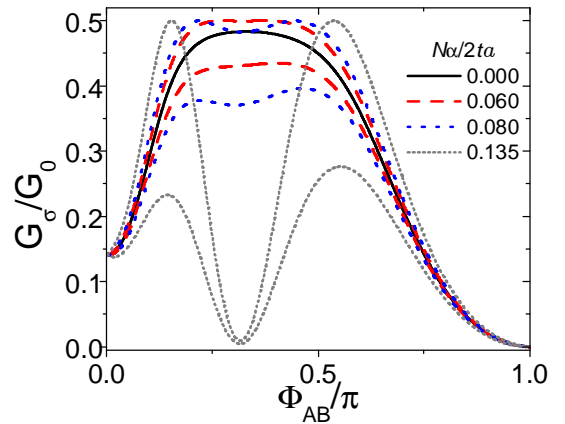}
\caption{Conductance for each spin as a function of flux for the
same parameters as in Fig. \protect\ref{rashba1} and different values of 
$\alpha$.}
\label{rashba2}
\end{figure}

Another important effect of the SOC in the Kondo regime is that it leads to
currents with significant spin polarization. If a spin $\sigma $ (up or
down) in the quantization direction $\vec{n}_{\theta }$ is injected in the
ring at the right lead ($\varphi =0$) it comes out at the left lead 
($\varphi =\pi $) with spin $\sigma $ in the direction 
$\vec{n}_{\theta}^{\prime }=(\sin {\theta },0,\cos {\theta })$ 
or vice versa \cite{note_homomorphism}.
The corresponding conductance $G_{\sigma }$ is spin dependent, 
as shown in Fig. \ref{rashba2}. 
The ratio of the conductances can reach a factor 2 or
larger with ideal $G_{\uparrow }$ ($G_{\downarrow }$) for flux 
$\Phi_{AB}=0.15\pi$ ($-0.15\pi $) and rather small $\alpha$ \cite{note_invariance}. 
For these values, the $z$ component of the quantization axis 
for any $\varphi$ is larger than 0.99 \cite{note_homomorphism}. 

In summary, we have presented an approach to calculate the conductance
through a ring of interacting QDs threaded by a magnetic flux and with 
spin-orbit coupling $\alpha $ in the Kondo, mixed-valence and non-magnetic regimes. 
The effects
of $\alpha $ are incorporated into Aharonov-Casher phases using a 
gauge transformation 
that leads
to the Hubbard Hamiltonian Eqs. (\ref{hub}) and (\ref{ac}). Using a method
based on a mapping of the relevant exact eigenstates of the ring onto an
effective multilevel Anderson impurity and with the use of a slave-boson
representation in the saddle-point approximation, we are able to describe
the properties of the ring connected to the leads. The method is valid for
small values of the coupling between rings and leads $V$ and small values of 
magnetic field $B$, 
such that the Zeeman energy is much less than $T_K$. 
When the ring
is in the Kondo regime, we obtain ideal conductance for $\alpha =0$ and
magnetic flux far from half a flux quantum, for which there is complete
destructive interference. The effect of a small non-vanishing $\alpha $ is to
produce a progressive destruction of the Kondo effect, decreasing the
conductance and leading to a strong spin dependence of it.
Extensions to include the Zeeman term or other interacting systems with
local interactions are straightforward. 

AAA wants to thank K. Hallberg, L. Arrachea and B. Normand for
useful discussions. We are partially supported by CONICET. This work was
sponsored by PIP 5254 of CONICET and PICT 03-13829 of ANPCyT. 
\bibliographystyle{apsrev}


\begin{thebibliography}{21}
\expandafter\ifx\csname natexlab\endcsname\relax\def\natexlab#1{#1}\fi
\expandafter\ifx\csname bibnamefont\endcsname\relax
  \def\bibnamefont#1{#1}\fi
\expandafter\ifx\csname bibfnamefont\endcsname\relax
  \def\bibfnamefont#1{#1}\fi
\expandafter\ifx\csname citenamefont\endcsname\relax
  \def\citenamefont#1{#1}\fi
\expandafter\ifx\csname url\endcsname\relax
  \def\url#1{\texttt{#1}}\fi
\expandafter\ifx\csname urlprefix\endcsname\relax\def\urlprefix{URL }\fi
\providecommand{\bibinfo}[2]{#2}
\providecommand{\eprint}[2][]{\url{#2}}

\bibitem[{\citenamefont{Berry}(1984)}]{berry84}
\bibinfo{author}{\bibfnamefont{M.}~\bibnamefont{Berry}},
  \bibinfo{journal}{Proc. R. Soc. Lond. A} \textbf{\bibinfo{volume}{392}},
  \bibinfo{pages}{45} (\bibinfo{year}{1984}).

\bibitem[{\citenamefont{Aharonov and Bohm}(1959)}]{aharonov&bohm59}
\bibinfo{author}{\bibfnamefont{Y.}~\bibnamefont{Aharonov}} \bibnamefont{and}
  \bibinfo{author}{\bibfnamefont{D.}~\bibnamefont{Bohm}},
  \bibinfo{journal}{Phys. Rev.} \textbf{\bibinfo{volume}{115}},
  \bibinfo{pages}{485} (\bibinfo{year}{1959}).

\bibitem[{\citenamefont{Ji et~al.}(2000)}]{ji00}
\bibinfo{author}{\bibfnamefont{Y.}~\bibnamefont{Ji}}
  \bibnamefont{\textit{et~al.}}, \bibinfo{journal}{Science} \textbf{\bibinfo{volume}{290}},
  \bibinfo{pages}{779} (\bibinfo{year}{2000}).

\bibitem[{\citenamefont{van~der Wiel et~al.}(2000)}]{wiel00}
\bibinfo{author}{\bibfnamefont{W.~G.} \bibnamefont{van~der Wiel}}
  \bibnamefont{\textit{et~al.}}, \bibinfo{journal}{Science} \textbf{\bibinfo{volume}{289}},
  \bibinfo{pages}{2105} (\bibinfo{year}{2000}).

\bibitem[{\citenamefont{Jiang et~al.}(2004)}]{jiang04}
\bibinfo{author}{\bibfnamefont{Z.-T.} \bibnamefont{Jiang}}
  \bibnamefont{\textit{et~al.}}, \bibinfo{journal}{Phys. Rev. Lett.}
  \textbf{\bibinfo{volume}{93}}, \bibinfo{pages}{076802}
  (\bibinfo{year}{2004});  
  \bibinfo{author}{\bibfnamefont{J.}~\bibnamefont{K{\"o}nig}},
   \bibinfo{author}{\bibfnamefont{Y.}~\bibnamefont{Gefen}}  \bibnamefont{and}
   \bibinfo{author}{\bibfnamefont{A.}~\bibnamefont{Silva}},
    \bibinfo{journal}{Phys. Rev. Lett.} \textbf{\bibinfo{volume}{94}}, \bibinfo{pages}{179701}
  (\bibinfo{year}{2005}); \bibinfo{author}{\bibfnamefont{Z.-T.} \bibnamefont{Jiang}}
  \bibnamefont{\textit{et~al.}}, \bibinfo{journal}{Phys. Rev. Lett.}
  \textbf{\bibinfo{volume}{94}}, \bibinfo{pages}{179702}
  (\bibinfo{year}{2005}) .

\bibitem[{\citenamefont{Meir and Wingreen}(1992)}]{meir92}
\bibinfo{author}{\bibfnamefont{Y.}~\bibnamefont{Meir}} \bibnamefont{and}
  \bibinfo{author}{\bibfnamefont{N.~S.} \bibnamefont{Wingreen}},
  \bibinfo{journal}{Phys. Rev. Lett.} \textbf{\bibinfo{volume}{68}},
  \bibinfo{pages}{2512} (\bibinfo{year}{1992}).

\bibitem[{\citenamefont{Jagla and Balseiro}(1993)}]{jagla93}
\bibinfo{author}{\bibfnamefont{E.~A.} \bibnamefont{Jagla}} \bibnamefont{and}
  \bibinfo{author}{\bibfnamefont{C.~A.} \bibnamefont{Balseiro}},
  \bibinfo{journal}{Phys. Rev. Lett.} \textbf{\bibinfo{volume}{70}},
  \bibinfo{pages}{639} (\bibinfo{year}{1993}).

\bibitem[{\citenamefont{Hallberg et~al.}(2004)}]{hallberg04}
\bibinfo{author}{\bibfnamefont{K.}~\bibnamefont{Hallberg}}
  \bibnamefont{\textit{et~al.}}, \bibinfo{journal}{Phys. Rev. Lett.}
  \textbf{\bibinfo{volume}{93}}, \bibinfo{pages}{067203}
  (\bibinfo{year}{2004}).

\bibitem[{\citenamefont{Aligia et~al.}(2004)}]{aligia04}
\bibinfo{author}{\bibfnamefont{A.~A.} \bibnamefont{Aligia}}
  \bibnamefont{\textit{et~al.}}, \bibinfo{journal}{Phys. Rev. Lett.}
  \textbf{\bibinfo{volume}{93}}, \bibinfo{pages}{076801}
  (\bibinfo{year}{2004}).

\bibitem[{\citenamefont{Pletyukhov et~al.}(2006)\citenamefont{Pletyukhov,
  Gritsev, and Pauget}}]{pletyukhov06}
\bibinfo{author}{\bibfnamefont{M.}~\bibnamefont{Pletyukhov}},
  \bibinfo{author}{\bibfnamefont{V.}~\bibnamefont{Gritsev}}, \bibnamefont{and}
  \bibinfo{author}{\bibfnamefont{N.}~\bibnamefont{Pauget}},
  \bibinfo{journal}{Phys. Rev. B} \textbf{\bibinfo{volume}{74}},
  \bibinfo{pages}{045301} (\bibinfo{year}{2006}).

\bibitem[{\citenamefont{Aharonov and Casher}(1984)}]{aharonov&casher84}
\bibinfo{author}{\bibfnamefont{Y.}~\bibnamefont{Aharonov}} \bibnamefont{and}
  \bibinfo{author}{\bibfnamefont{A.}~\bibnamefont{Casher}},
  \bibinfo{journal}{Phys. Rev. Lett.} \textbf{\bibinfo{volume}{53}},
  \bibinfo{pages}{319} (\bibinfo{year}{1984}).

\bibitem[{\citenamefont{K{\"o}nig et~al.}(2006)}]{konig06}
\bibinfo{author}{\bibfnamefont{M.}~\bibnamefont{K{\"o}nig}}
  \bibnamefont{\textit{et~al.}}, \bibinfo{journal}{Phys. Rev. Lett.}
  \textbf{\bibinfo{volume}{96}}, \bibinfo{pages}{076804}
  (\bibinfo{year}{2006}).

\bibitem[{\citenamefont{Bergsten et~al.}(2006)}]{bergsten06}
\bibinfo{author}{\bibfnamefont{T.}~\bibnamefont{Bergsten}}
  \bibnamefont{\textit{et~al.}}, \bibinfo{journal}{Phys. Rev. Lett.}
  \textbf{\bibinfo{volume}{97}}, \bibinfo{pages}{196803}
  (\bibinfo{year}{2006}).

\bibitem[{\citenamefont{Moln\'ar et~al.}(2004)\citenamefont{Moln\'ar, Peeters,
  and Vasilopoulos}}]{molnar04}
\bibinfo{author}{\bibfnamefont{B.}~\bibnamefont{Moln\'ar}},
  \bibinfo{author}{\bibfnamefont{F.~M.} \bibnamefont{Peeters}},
  \bibnamefont{and}
  \bibinfo{author}{\bibfnamefont{P.}~\bibnamefont{Vasilopoulos}},
  \bibinfo{journal}{Phys. Rev. B} \textbf{\bibinfo{volume}{69}},
  \bibinfo{pages}{155335} (\bibinfo{year}{2004}); \bibinfo{author}{\bibfnamefont{D.}~\bibnamefont{Frustaglia}} \bibnamefont{and}
  \bibinfo{author}{\bibfnamefont{K.}~\bibnamefont{Richter}},
  \bibinfo{journal}{Phys. Rev. B} \textbf{\bibinfo{volume}{69}},
  \bibinfo{pages}{235310} (\bibinfo{year}{2004}); \bibinfo{author}{\bibfnamefont{G.~S.} \bibnamefont{Lozano}} \bibnamefont{and}
  \bibinfo{author}{\bibfnamefont{M.~J.} \bibnamefont{S\'anchez}},
  \bibinfo{journal}{Phys. Rev. B} \textbf{\bibinfo{volume}{72}},
  \bibinfo{pages}{205315} (\bibinfo{year}{2005}); \bibinfo{author}{\bibfnamefont{S.}~\bibnamefont{Souma}} \bibnamefont{and}
  \bibinfo{author}{\bibfnamefont{B.~K.} \bibnamefont{Nikoli\'c}},
  \bibinfo{journal}{Phys. Rev. Lett.} \textbf{\bibinfo{volume}{94}},
  \bibinfo{pages}{106602} (\bibinfo{year}{2005}).


\bibitem[{\citenamefont{Meijer et~al.}(2002)\citenamefont{Meijer, Morpurgo and
  Klapwijk}}]{meijer02}
\bibinfo{author}{\bibfnamefont{F.~E.} \bibnamefont{Meijer}},
  \bibinfo{author}{\bibfnamefont{A.~F.} \bibnamefont{Morpurgo}}
  \bibnamefont{and} \bibinfo{author}{\bibfnamefont{T.~M.}
  \bibnamefont{Klapwijk}}, \bibinfo{journal}{Phys. Rev. B}
  \textbf{\bibinfo{volume}{66}}, \bibinfo{pages}{033107}
  (\bibinfo{year}{2002}).

\bibitem[{\citenamefont{Meir et~al.}(1989)\citenamefont{Meir, Gefen, and
  Entin-Wohlman}}]{meir89}
\bibinfo{author}{\bibfnamefont{Y.}~\bibnamefont{Meir}},
  \bibinfo{author}{\bibfnamefont{Y.}~\bibnamefont{Gefen}}, \bibnamefont{and}
  \bibinfo{author}{\bibfnamefont{O.}~\bibnamefont{Entin-Wohlman}},
  \bibinfo{journal}{Phys. Rev. Lett.} \textbf{\bibinfo{volume}{63}},
  \bibinfo{pages}{798} (\bibinfo{year}{1989}).

\bibitem[{\citenamefont{Gritsev et~al.}(2005)}]{gritsev05}
\bibinfo{author}{\bibfnamefont{V.}~\bibnamefont{Gritsev}} \bibnamefont{\textit{et~al.}},
  \bibinfo{journal}{Phys. Rev. Lett.} \textbf{\bibinfo{volume}{94}},
  \bibinfo{pages}{137207} (\bibinfo{year}{2005}).

\bibitem[{not({\natexlab{a}})}]{note_homomorphism}
\bibinfo{note}{Using the well known SU(2) $\longrightarrow $ SO(3)
  homomorphism, it can be shown that the quantization axis at an angle $\varphi
  $ is obtained simply rotating $\vec{n}_{\theta }$ by $\varphi $ around the
  $z$ axis.}

\bibitem[{\citenamefont{Lobos and Aligia}(2006)}]{lobos06}
\bibinfo{author}{\bibfnamefont{A.~M.} \bibnamefont{Lobos}} \bibnamefont{and}
  \bibinfo{author}{\bibfnamefont{A.~A.} \bibnamefont{Aligia}},
  \bibinfo{journal}{Phys. Rev. B} \textbf{\bibinfo{volume}{74}},
  \bibinfo{pages}{165417} (\bibinfo{year}{2006}).

\bibitem[{\citenamefont{Coleman and Andrei}(1986)}]{coleman86}
\bibinfo{author}{\bibfnamefont{P.}~\bibnamefont{Coleman}} \bibnamefont{and}
  \bibinfo{author}{\bibfnamefont{N.}~\bibnamefont{Andrei}},
  \bibinfo{journal}{J. Phys. C} \textbf{\bibinfo{volume}{19}},
  \bibinfo{pages}{3211} (\bibinfo{year}{1986}).

\bibitem[{not({\natexlab{b}})}]{note_invariance}
\bibinfo{note}{The Hamiltonian is invariant under time reversal and change of
  sign of $\Phi _{AB}=2(\pi r)^2 B /\phi _{0}$.}

\end{thebibliography}


\end{document}